\documentclass[11pt, reqno]{amsart}

\usepackage[body={6.3in,9.3in}]{geometry}
\usepackage[ansinew]{inputenc}

\usepackage{textcomp}
\usepackage{latexsym}
\usepackage[usenames]{color}
\usepackage{amsmath,amsfonts,amssymb,amsthm}
\usepackage{graphicx}
\usepackage{longtable}
\usepackage{bbm}

\usepackage{setspace}
\usepackage{array}

\def\qed{\hfill{\raggedleft{\hbox{$\Box$}}} \smallskip}

\def\S{\mathbb{S}}
\def\R{\mathbb{R}}

\def\X{\mathcal{X}}
\def\A{\mathcal{A}}
\def\ST{\mathcal{ST}}

\DeclareMathOperator{\trop}{trop}
\DeclareMathOperator{\adj}{adj}
\DeclareMathOperator{\tr}{tr}

\newcommand{\ds}{\displaystyle}
\newcommand{\vect}[1]{\overrightarrow{#1}}
\renewcommand{\iff}{\Leftrightarrow}
\newcommand{\imply}{\Rightarrow}

\theoremstyle{plain} \newtheorem{lem}{Lemma}[section]
\theoremstyle{plain} 
\theoremstyle{plain} \newtheorem{thm}[lem]{Theorem}
\theoremstyle{plain} \newtheorem{cor}[lem]{Corollary}
\theoremstyle{plain} \newtheorem{conj}[lem]{Conjecture}
\theoremstyle{definition} 
\theoremstyle{definition}
\theoremstyle{definition} 
\theoremstyle{definition}\newtheorem{ex}[lem]{Example}

\newenvironment{Proof}{ {\bf Proof.}\ 
                           \begingroup }
                           {\endgroup ~ \qed \vskip10pt \noindent}        

\newlength\savedwidth

\begin{document}

\title{Pairwise ranking: choice of method can produce arbitrarily different rank order}
\author{Ngoc Mai Tran}
\address{ Department of Statistics,University of California at Berkeley,  Berkeley, CA 94720-3860 U.S.A.}
\email{tran@stat.berkeley.edu}
\begin{abstract}
 We examine three methods for ranking by pairwise comparison: Principal Eigenvector, HodgeRank and Tropical Eigenvector. It is shown that the choice of method can produce arbitrarily different rank order.To be precise, for any two of the three methods, and for any pair of rankings of at least four items, there exists a comparison matrix for the items such that the rankings found by the two methods are the prescribed ones. We discuss the implications of this result in practice, study the geometry of the methods, and state some open problems.
\end{abstract}
\maketitle

\section{Introduction}
The problem of ranking appears in various contexts and forms. The choice of ranking method is often dictated by the format of the input data. Here we consider the following pairwise comparison model: suppose there are $n$ items to be ranked. Let $A$ denote the $n \times n$ pairwise comparison matrix, where $A_{ij}$ measures how much $i$ dominates over $j$. These numbers are usually deduced from an experiment, opinion poll or voting behavior. The goal is to produce a cardinal ranking, one in which items are ranked based on a score vector $a \in \R^n$. 

Frequently in practice the comparisons $A_{ij}$ are recorded on either a multiplicative scale $(A_{ij} = 1/A_{ji}, A_{ij} > 0)$, or an additive scale ($A_{ij} = -A_{ji},A_{ij} \in \R$). We shall use the term \textit{multiplicative paired comparison matrix} (MPCM) and \textit{additive paired comparison matrix} (APCM) to distinguish them. Both classes of matrices are found in various applications. The APCMs are sometimes derived from large-scale modern internet and e-commerce databases as a mean to give information on the ranking of items \cite{lekheng}. Many other ranking models can also be casted in this framework, for example, the APCMs with entries restricted to $\{\pm 1\}$ are precisely the matrices appear in the standard scenario of binary pairwise comparisons \cite{lekheng}. On the other hand, MPCMs, also called \textit{symmetrically reciprocal matrices} \cite{elsner, farkas}, are widely used in business and related fields. They were first suggested by Saaty \cite{fuzzy}, who proposed the Principal Eigenvector method for obtaining a score vector from these matrices, and used them in the Analytic Hierarchy Process for multi-criteria decision making. 

Throughout this paper we shall reserve $A$ to indicate an APCM, and $X$ to indicate an MPCM. Given the natural one-to-one correspondence between these two classes of matrices via taking elementwise log/exponential, we shall state definitions in terms of an APCM $A$ only, with the corresponding definition for an MPCM $\left(X_{ij}\right) = \left(\exp_b{(A_{ij})}\right)$ implied, where $b \neq 0$ can be any choice of exponent base. Following \cite{saari}, we say that a paired comparison matrix $A$ is \emph{strongly transitive} if 
$$A_{ik} = A_{ij} + A_{jk} \mbox{ for all } i, j, k,$$ 
and say that $A$ is \emph{transitive} if 
$$A_{ij},A_{jk} > 0 \imply A_{ik} > 0.$$
Note that in either cases the task of retrieving an ordinal ranking is trivial. Furthermore, strong transitivity implies the existence of a unique cardinal ranking up to constants \cite{fuzzy}:
\begin{equation} 
 A \mbox{ is strongly transitive} \iff A_{ij} = a_i - a_j \label{eqn:st}
\end{equation}
for some $a \in \R^n$ unique up to an additive constant. 
\vskip12pt \noindent
Here $a_i$ has the natural interpretation of being the score assigned to the $i^{th}$ item relative to some reference point, corresponding to the additive constant. In practice $A$ is likely to be corrupted by noise, or obtained from averaging the opinion of multiple individuals, and therefore may not be strongly transitive nor transitive. Equation (\ref{eqn:st}) shows that finding a score vector is equivalent to approximating the observed matrix by a strongly transitive matrix. One can formulate this as follows.
\vskip12pt \noindent
\textbf{Problem.} Suppose that there is an unknown true score vector $s \in \R^n$ which induces a strongly transitive additive comparison matrix $S$, and that we observe a noisy APCM $A$, where 
\begin{equation} A = S + \epsilon \label{eqn:x.eps} \end{equation}
for some noise matrix $\epsilon$. Can we recover the true score vector $s$ up to an additive constant, or at least up to the same induced ranking?
\vskip12pt \noindent
A similar formulation holds for MPCMs. A number of methods for estimating $s$ have been proposed in the literature, some are designed specifically for MPCMs \cite{elsner, fuzzy, crawford}, while others are for APCMs \cite{lekheng}. In general these methods can produce different rankings, and the choice of method remains a topic of debate. A number of interesting questions arise: when and how often do these methods produce different rankings? Can they produce `very different' rankings? Does the format of the data (APCM or MPCM) matter? If one wants to convert from one type to another, does the choice of base $b$ in defining the log/exponential map matter?
\vskip12pt 
In this paper, we aim to answer the above questions for three specific methods commonly used in practice: Principal Eigenvector, HodgeRank and Tropical Eigenvector. Some properties of the first two have been studied and compared in the literature \cite{lekheng, fuzzy, crawford, dong}. Our main result is the following.
\begin{thm}{}\label{thm}
For $n \geq 4$, for any two of the above three methods, given any pairs of rank order $(\sigma_1,\sigma_2)$, there exists a paired comparison matrix $A$ such that the first method applied to $A$ gives the ranking $\sigma_1$, and the second method applied to $A$ gives the ranking $\sigma_2$ on the $n$ items. 
\end{thm}
This result is somewhat surprising and may remind the reader of Arrow's Impossibility Theorem in social choice theory. In particular, it implies that these methods are fundamentally different, and the choice of methods in practice deserves further study. 
\vskip12pt
The paper is organized as follows: in Section \ref{sec:intro}, we describe the three methods, their known properties, and their connections. We state some lemmas concerning their behavior under permutation of items, scaling and perturbation of the matrix $A$. These will be used to prove Theorem \ref{thm}, but they are also interesting in their own rights. In Section \ref{sec:geom}, we study the geometry behind our theorem, which is revealed to be strongly connected to the geometry of the Tropical Eigenvector. We compute an explicit formula for the Tropical Eigenvector when $n = 4$. In Section \ref{sec:proof}, we prove Theorem ~\ref{thm} and discuss the problem of realizability for ranking triples. Section \ref{sec:discuss} describes some open problems.


\section{The three methods}\label{sec:intro}
\textbf{Notations.} Let $\mathcal{A}$ denote the set of APCMs, which are $n \times n$ skew-symmetric matrices, let $\ST \subset \A$ be the subspace of strongly transitive APCMs. Let $\mathcal{X}$ denote the set of MPCMs. For a matrix $X$, let $v(X), h(X), m(X)$ denote the principal eigenvector, HodgeRank vector, and tropical eigenvector of $X$, respectively. These will be defined below. Let $X^{(k)}$ denotes the $k^{th}$ Hadamard power ($X^{(k)}_{ij} = X_{ij}^k$). For two matrices $X$ and $Y$, let $X \circ Y$ denote their Hadamard product: $(X\circ Y)_{ij} = X_{ij}Y_{ij}$. As an abuse of notation we shall also use $A$ to indicate the vector $(A_{ij}: i < j)$, since it should be clear from context. By the $L_\infty$-norm of $A$ we mean $\|A\|_\infty = \max_{i,j}|A_{ij}|$, and the $L_2$-norm of $A$ is $\|A\|_2 = \sqrt{\sum_{i,j}A_{ij}^2}$. For $M \in \ST$, let $\vect{M} := -(M_{12},\ldots,M_{1n})$ be the last $n-1$ coordinates of the score vector $m$ of $M$, normalized so $m_1 = 0$. For any set $S \subset \ST$, define $\vect{S} = \{\vect{M}, M \in S\}$. For $A \in \A$, \emph{the graph of $A$} is the directed graph where $A_{ij}$ is the weight of the edge from $i$ to $j$.  For a cycle $\pi: i_1\rightarrow i_2 \rightarrow \ldots \rightarrow i_k \rightarrow i_{k+1} = i_1$, the cycle value $f(\pi)$ is defined to be the sum of the corresponding edges: $f(\pi) := \sum_j A_{i_ji_{j+1}}.$

As we shall prove (Lemma \ref{lem:list.prop}, part 1), all three methods are invariant under relabeling of the items. Therefore, we shall think of a ranking as a permutation, that is, an element of the symmetric group on $n$ letters $\S^n$. For a generic vector $w \in \R^n$, let $r(w) \in \S_n$ denote its induced ranking. Say that $A$ \emph{induces a ranking triple} $(\sigma_1,\sigma_2,\sigma_3)$ for $(v,h,m)$ if $r(v) = \sigma_1, r(h) = \sigma_2, r(m) = \sigma_3$. Say that a ranking triple $\sigma = (\sigma_1,\sigma_2,\sigma_3) \in \S_n^3$ is \emph{realizable for $(v,h,m)$} if there exists a matrix $A$ which induces that ranking triple. Definitions for the ranking pairs follow similarly. 
\vskip12pt \noindent
\subsection{Principal Eigenvector}
For $X \in \X$, note that $X_{ij} > 0$ for all $i,j$, therefore its principal eigenvector $v(X)$ is real and unique by the Perron-Frobenius theorem, and is used as an estimator for the score vector in this method.

Since its existence only relies on the non-negativity of the matrix entries, the Principal Eigenvector method is also used in ranking contexts other than paired comparisons. For example, in Google's PageRank algorithm, one starts with a matrix that essentially consists of transition probabilities between webpages of a random web surfer and uses its principal eigenvector for ranking webpages. The specific application of Principal Eigenvector to MPCMs first appeared in the Analytic Hierarchy Process \cite{fuzzy}, where it is utilized to obtain the score vector at intermediate steps in the hierarchy. 

The Principal Eigenvector has a Bayesian interpretation: given a prior score vector $p \in \R^n_+$, we can think of $Xp$ as the posterior score, since 
\[ (Xp)_i = \sum_j X_{ij} p_j =  \mbox{sum over all $j$ of how much $i$ wins $j$, weighted by our prior $p_j$}. \] 
Then the Principal Eigenvector $v(X)$ is the Bayesian solution to the ranking problem, since, as an element of the projective space $\mathbb{PR}^{n-1}$ it is invariant under further updates by $X$.
\vskip12pt \noindent
\subsection{HodgeRank}
Here one equips the linear space $\mathcal{A}$ with the usual matrix inner product. The HodgeRank vector $h(A)$ is the score vector of the $L_2$ projection of $A$ onto the subspace $\ST$. That is, it minimizes the $L_2$-norm of the error $\epsilon$ in Equation ~ (\ref{eqn:x.eps}). It admits a simple formula.
\begin{lem}{\cite[Theorem 3]{crawford}}\label{lem:h} 
Let $A \in \mathcal{A}$. Then, up to an additive constant, the HodgeRank vector $h(A)$ is given by the normalized row sum of $A$.
\[ h(A) = \frac{1}{n} A \cdot \mathbf{1} \]
where $\mathbf{1}$ is $n \times 1$ column vector consists of 1's.
\end{lem}
\vskip12pt \noindent
\textbf{Proof.} Let $H \in \ST$ be the strongly transitive matrix defined by $h$. It is sufficient to show that $\tr((A-H)^TW) = 0$ for all $W \in \ST$. Let $w$ be a score vector of $W$. Let $r_i, c_i$ denotes the $i^{th}$ row and column sum of $A$, respectively. Note that $c_i = -r_i$ since $A$ is skew-symmetric. Then
\begin{align*}
\tr((A-H)^TW) &= \sum_i\sum_j\left(A_{ij} - \frac{r_i}{n} + \frac{r_j}{n}\right)(w_i - w_j) \\
&= \sum_iw_i(r_i - r_i) - \sum_jw_j(c_j + r_j) = 0. \hspace{4em} \qed
\end{align*}
Lemma \ref{lem:h} indicates that one can also apply HodgeRank to an MPCM, in which case the $i^{th}$ entry of the score vector is the geometric mean of the $i^{th}$ row. This was first noticed by Crawford \cite{crawford}, who proposed the name `row geometric mean', or `log least square estimator'. The name `HodgeRank' was proposed by Jiang {\em et al.} \cite{lekheng} who approached the problem using combinatorial graph theory. In this case the entries $A_{ij}$ are viewed as edge flows from node $i$ to node $j$, and the score vector is the potential of the nodes. Entries of $h(A)$ can be interpreted as the average outflow from each node, which is an intuitive measure of its potential. This method has a number of appealing properties both computationally and mathematically, as studied in details in \cite{lekheng}. For example, it is the maximum likelihood estimator for $s$ under certain classes of statistical models on the error term $\epsilon$. The special case of multiple raters with i.i.d normal error was discovered earlier by Crawford \cite{crawford}. 
\vskip12pt \noindent
\subsection{Tropical Eigenvector}
This ranking method was suggested by Elsner and van den Driessche \cite{elsner}. Here, one considers $A$ under the max-plus algebra $(\mathbb{R}, \oplus, \odot)$, where $x\oplus y = \max{\{x,y\}}, x\odot y = x + y$. If a pair $(\lambda_{max}, m) \in \R \times \R^n$ satisfy
\[ A \odot m = \lambda_{max} \odot m \]
where the matrix multiplication is performed in the max-plus algebra, then it is called a \emph{tropical eigenvalue-eigenvector} pair of $A$. It is known (see, for example, \cite{bernd.trop, butkovic}) that if $A$ is irreducible, then $\lambda_{max}$ is unique, and it is equal to the maximal normalized cycle length of the graph of $A$, that is,
\[ \lambda_{max} = \max_{\{i_1,\ldots, i_k\} \subseteq \{1,\ldots,n\}}\frac{1}{k}\left(A_{i_1i_2} + A_{i_2i_3}+\ldots+ A_{i_{k-1}i_k}+A_{i_ki_1}\right). \]
Note that this value can be attained by more than one cycle in $A$. Such cycles are called \textit{critical cycles}, and their vertices are called \textit{critical vertices} \cite{elsner}. One can also apply this method to an MPCM $X$ by considering it under the tropical max-times algebra $(\R, \oplus, \cdot): x\oplus y = \max{\{x,y\}}, x \cdot y = xy$ as in \cite{elsner}. Then, the tropical max-times eigenvector $v_{max}(X)_i = \exp_bm(A)_i$ serves as an estimator for the true score vector. It is easy to see that different choices of the exponent base $b$ amounts to different scalings of $v_{max}$, and therefore preserves the induced ranking.

 In the Tropical Eigenvector method, one uses the tropical eigenvector $m(A)$ as an estimator for the true score vector $s$. For all matrices $A \in \A$, $\lambda_{max}$ is always unique, and $m(A)$ is unique if and only if the critical cycles of $A$ are not disjoint \cite{elsner,bapat}. This is a very mild condition: the set of matrices with non-unique critical cycles is a finite union of disconnected polyhedral sets of measure 0 under the Lebesgue measure on $\A$. For `general' $A$ (that is, for most reasonable, generic distributions on the noise $\epsilon$), $\lambda_{max}$ is almost surely attained by only one cycle, and hence the uniqueness (up to tropical scaling) of $m$ is guaranteed almost surely. 

\subsection{Properties of the three methods}
From now on, we shall assume that the matrix $A$ under consideration has a unique tropical eigenvector. In this section, we state a collection of results concerning the behavior of these estimators under permutation of the labeling of the items, scaling and perturbation of the matrix $A$. The proof relies on some properties of the tropical eigenvector, and is therefore deferred to Section \ref{subsec:proof.lem}. Here we shall focus on interpreting its implications. In particular, we address the issue of choice of exponent $b$ when converting between APCMs and MPCMs. 
\begin{lem}{}\label{lem:list.prop} 
\textbf{(1)\hspace{1em}} {All three methods are invariant under relabeling of the items.}  \\
That is, for $\tau \in \S_n$, let $[\tau]$ denotes the permutation matrix that maps the vector $[1 \, 2\, \ldots \, n]$ to $[\tau(1)\, \tau(2) \, \ldots \, \tau(n)]$. If $A$ induces the ranking triple $(\sigma_1,\sigma_2,\sigma_3)$, then $Y := [\tau] A [\tau^{-1}]$ induces the ranking triple $(\tau \sigma_1,\tau \sigma_2, \tau\sigma_3)$.
\vskip12pt \noindent
\textbf{(2)\hspace{1em}} {HodgeRank and Tropical Eigenvector are invariant under multiplication of $A$ by scalars, or taking Hadamard powers of $X$. Principal Eigenvector does not have this property.} That is, for any constants $c \neq 0, k > 0$, for $A \in \A$ and $X \in \X$, we have
\begin{align*}
h(c \cdot A) = h(A) & \hspace{1em} h(X^{(k)}) = h(X)^k \\
m(c \cdot A) = c \cdot m(A) & \hspace{1em}  m(X^{(k)}) = m(X)^k
\end{align*}
\textbf{(3)\hspace{1em}} {HodgeRank is additive in $\A$ and Hadamard multiplicative in $\X$. Principal Eigenvector and Tropical Eigenvector are Hadamard multiplicative in $\X$ with strongly transitive matrices. } That is, if $A, A' \in \A$ and $X, X' \in \X$, then for all $i$,
$$h(A + A')_i = h(A)_i + h(A')_i, \hspace{1em} h(X \circ X')_i = h(X)_ih(X')_i$$
If in addition $A$ and $X$ are strongly transitive, then for all $i$,
$$m(A + A')_i = m(A)_i + m(A')_i, \hspace{1em} m(X \circ X')_i = m(X)_im(X')_i,$$
and \hspace{8em} $v(X \circ X')_i = v(X)_iv(X')_i.$
\end{lem}
\vskip12pt \noindent
\textbf{Implications on conversion between APCMs and MPCMs} \\
For fixed $b$, consider the isomorphism from $\A$ to $\X$ defined via 
\[ A \mapsto X := (X_{ij}) = (\exp_b(A_{ij})).\] 
Note that $\exp_b(A_{ij}) = \exp(\log(b) \cdot A_{ij})$. Since HodgeRank and Tropical Eigenvector are invariant under scaling, the choice of $b$ does not changes the induced ranking. Multiplying $A_{ij}$ by a constant corresponds a change of measurement scale, and therefore it is desirable that the ranking does not change when one changes measurement unit. Hence, in considering the induced ranking of HodgeRank and Tropical Eigenvector, one can work entirely in $\A$ and utilize its inner product space structure, and then lift to $\X$ via taking element-wise exponential using any base, and vice versa.

However, one cannot do the same with Principal Eigenvector, since Lemma \ref{lem:list.prop} part 2 indicates that the choice of $b$ affects the ranking. This does not imply that the method depends on the measurement unit: in fact, since the entries of $X$ measure the \emph{ratio} between two scores, they are by default unit-free. The lack of scale invariance of Principal Eigenvector simply means that the method is not designed for additive paired comparison data. In general, if a ranking method is invariant under all monotone transformations of the entries of $A$ (or $X$), then it must be a method that only take into account the signs of the entries of $A$, that is, $A_{ij} \in \{\pm 1, 0\}$. In this case, it only makes sense to ask for an ordinal and not cardinal ranking. Therefore, in working with cardinal ranking, any method lacks the invariance property under \emph{some} monotone transformations of the data. The bottom line is, for Principal Eigenvector, one should always work with MPCMs, while HodgeRank and Tropical Eigenvector can be applied to both MPCMs and APCMs. However, as we shall see in Section \ref{sec:geom}, there is a version of Principal Eigenvector on APCMs, and this turns out to be the Tropical Eigenvector.
\section{Geometry of the methods}\label{sec:geom}
In the first part of this section, we shall see that Tropical Eigenvector on $\X$ can be interpreted as the scale-invariant limit of Principal Eigenvector, and Tropical Eigenvector on $\A$ has the geometric interpretation of being a special point on the set of $L_\infty$-minimizers of $\epsilon$ from $A$ to $\ST$, thus is related to HodgeRank, which is the $L_2$-minimizer of $\epsilon$. Therefore, the Tropical Eigenvector serves as a bridge connection the other two methods. The rest of the section will be devoted to looking at the geometry of Tropical Eigenvector in general and for $n = 4$ in particular.
\subsection{Tropical as Principal Eigenvector on $\A$}
Suppose one wants to apply Principal Eigenvector to APCMs. Then one seeks for a scale-invariant version of this method, and interestingly, the Tropical Eigenvector is the answer. This interpretation is based on the following result of Gaubert {\em et al.} \cite{gaubert}.
\begin{cor}{\cite{gaubert}}\label{cor:gaubert}
If the max-times tropical eigenvector $v_{max}(X)$ of $X \in \X$ is unique, then it can be obtained as the limit of the principal eigenvector of Hadamard powers of $X$. That is,
\[ v_{max} = \lim_{k\rightarrow\infty} (v(X^{(k)}))^{1/k} \]
where $v(X^{(k)})$ denote the principal eigenvector of $X^{(k)}$.
\end{cor}
One could interpret this in the context of rank aggregation as follows: suppose there are $k$ raters who produce identical pairwise comparison matrices 
\[ A = A_1 = \ldots = A_k.\] 
If one uses Principal Eigenvector to obtain an overall ranking by averaging the score of the total preferences (aggregated by taking Hadamard product of the individual matrices), then one obtains the tropical eigenvector in the limit as the number of raters $k \rightarrow \infty$. This gives Tropical Eigenvector the interpretation of being `Principal Eigenvector on APCMs'. 
\subsection{Tropical as $L_\infty$ minimization on $\A$}
Let $D_A$ be the set of $L_\infty$-minimizers from $A$ to $\ST$, let $\trop(A) = \{m \in \R^n: A \odot m = \lambda_{max}(A) \odot m\}$ denote its tropical eigenspace. A restatement of Theorem 2 in \cite{elsner10} gives the following.
\begin{cor} \label{cor:linfty} \cite{elsner10}
The max-plus eigenvalue $\lambda_{max}(A)$ is the $L_\infty$ distance between $A$ and the subspace $\ST$, and the tropical eigenspace of $A$ is a subset of the score vectors of $L_\infty$-minimizers from $A$ to $\ST$. That is,
\[ \trop(A) \subseteq \vect{D_A}. \]
In particular, one tropical eigenvector of $A$ is given by
\begin{equation} 
m_j = \min\{w_j: w \in \vect{D_A}\} = \max\{W_{1j}: W \in D_A\}. \label{eqn:formula.m}
\end{equation}
In other words, the upper right corner with respect to the standard basis on $\A$ of the polytope $D_A$ is always a tropical eigenvector. 
\end{cor}
Corollary \ref{cor:linfty} makes the link between HodgeRank and tropical max-plus eigenvector clear: the former is the $L_2$-minimizer, while the later is the upper right corner in the set of $L_\infty$-minimizers from $A$ to $\ST$ when it is unique.
A complete characterization of cases where $\trop(A) = \vect{D_A}$ is not yet known. One such instance is when $D_A$ is a point, and this occurs if and only if the unique critical cycle of $A$ is an $n$-cycle \cite{elsner10}. 
\subsection{Geometry of the Tropical Eigenvector}\label{sec:geomtrop}
Both HodgeRank and Tropical Eigenvector rely on the  orthogonal decomposition $\A = \ST \oplus \ST^\perp$. Therefore, in studying the geometry of these methods, it is beneficial to work with a basis of $\A$ which reflects this decomposition. Saari \cite{saari} proposed a natural basis for $\ST$ and identified a spanning set for $\ST^\perp$. Here, we shall extend his work by showing that there is also a natural choice of basis for $\ST^\perp$. The following results were stated in \cite{saari} using different terminologies.
\begin{lem}{\cite{saari}}\label{lem:saari}
Regard $\A$ as $\R^{n \choose 2}$, so $A = (A_{ij}, i > j) \in \A$ is a vector. \\
\textbf{A basis for $\ST$}: For $i \in \{1,\ldots,n\}$, define $t_i \in \{\pm 1, 0\}^{n\choose 2}$ to be the vector such that 
$$\langle t_i, A \rangle = \mbox{the }i^{th}\mbox{ row sum of $A$ viewed as a skew-symmetric matrix},$$ 
where $\langle \cdot, \cdot \rangle$ is the usual inner product in $\R^{n \choose 2}$.\\ 
Then any subset of $n-1$ vectors from the set $\{t_1,\ldots,t_n\}$ defines a basis for $\ST$.
\vskip12pt \noindent
\textbf{A spanning set for $\ST^\perp$}: 
Fix a vector $A \in \A$. For each cycle $\pi$ in the graph of $A$, define $s(\pi) \in \{\pm1, 0\}^{n \choose 2}$ to be the vector such that
\[ f(\pi) = \langle s(\pi), A\rangle. \]
Then the set $S$ consisting of all vectors $s(\pi)$, where $\pi$ is a cycle in the graph of $A$, spans $\ST^\perp$.
\end{lem}
\vskip12pt \noindent
The cardinality of $S$, which is the number of $k$-cycles for $3 \leq k \leq n$ is much greater than ${n-1 \choose 2}$, the dimension of $\ST^\perp$. Clearly any linearly independent subset of size ${n-1 \choose 2}$ would be a basis. We note that there is a natural choice: namely, the set of 3-cycles involving a fixed vertex $v$.
\begin{lem}{A basis for $\ST^\perp$}\label{lem:basis.for.m.perp}
Fix a vertex $v$. Let
\[ S_v := \{s(\pi) \in \{\pm1, 0\}^{n \choose 2}: \pi \mbox{ is a 3-cycle with } \pi(1) = v\} \]
Then $S_v$ has $2 \times {n-1 \choose 2}$ vectors which occur in pairs of the form $(s(\pi),-s(\pi))$. Pick one representative for each pair. Then the resulting set is a basis of $\ST^\perp$.
\end{lem}
\vskip12pt \noindent
\begin{Proof} The cardinality of $S_v$ is $2 \times {n-1 \choose 2}$ since there are two choices for the other two nodes to complete an undirected $3$-cycle. Note that any $k$-cycle $\pi$ involving the node $v$ can be graphically represented by a planar $k$-gon, which we shall call the \textit{graph of $\pi$}. Since $A$ as a matrix is skew-symmetric, 
\[ s(\pi) = \sum_{j=1}^{k-2}s(\tau_j) \]
where $\{\tau_j \in S_v: j = 1,\ldots,k-2\}$ is a triangulation of the graph of $\pi$. It follows that $S_v$ spans the set of all $k$-cycles involving $v$. On the other hand, the graph of $\pi$ can also be triangulated by fixing another vertex $v'$. Since the choice of cycle is arbitrary, it follows that any other 3-cycle involving $v'$ is also in the span of $S_v$. An inductive argument shows that $S_v$ spans $S$, completing the proof.
\end{Proof}
The following corollary comes from a direct computation and gives the geometric interpretation of the tropical max-plus eigenvector of $A$.
\begin{cor}\label{cor:zonotope}
Fix $A \in \A$, viewing it as a vector in $\R^{n \choose 2}$. The projection of the standard cube centered at $A$ onto $\ST$ is the $(n-1)$-permutahedron, scaled by $1/n$ and centered at $h(A)$. Its dual zonotope $Z_\perp$, which is the projection of the standard cube centered $A$ onto $\ST^\perp$, is a convex symmetric polytope centered at $A - P(A)$, where $P(A)$ is the orthogonal projection of $A$ onto $\ST$. The distance in $\ST^\perp$ from the center of $Z_\perp$ to $0$ is $\lambda_{max}(A)$. Furthermore, the facet(s) of $\lambda_{max}(A) \cdot Z_\perp$ which contain $0$ are precisely the critical cycle(s) of $A$. 
\end{cor}
The zonotope $Z_\perp$ is the cographic zonotope associated with the
   complete graph on $n$ nodes (see, for example, \cite{ziegler}). It exhibits much symmetry, since actions of $\S_n$ on the labeling of the items correspond to rotations of $Z_\perp$. While the corollary applies independent of the choice of coordinate on $\ST^\perp$, in visualizing $Z_\perp$ some basis may expose the symmetry induced by the actions of $\S_n$ better than others. This is demonstrated in the following example.
\begin{ex}
Let $n = 4$. Here $\dim(\ST) = \dim(\ST^\perp) = 3$. We shall equip $\ST^\perp$ with a basis consisting of $4$-cycles. There are six 4-cycles in $\S_4$ and they occur in pairs of the form $(\sigma, \sigma^{-1})$. Choose a representative for each pair. Then the resulting set $S_\pi$ consists of 3 independent vectors in $\R^6$, and they form a basis for $\ST^\perp$. For concreteness, we shall choose $S_\pi$ to be the set $\{f_1,f_2,f_3\}$, with
\[ f_1 = s((1 \, 2 \, 3 \, 4)), \hspace{1em} f_2 = s((1 \, 3 \, 4 \, 2)), \hspace{1em} f_3 = s((1\, 4\, 2\, 3)), \] 
where $s(\pi)$ of a cycle $\pi$ is the unique vector defined in Lemma \ref{lem:saari}. Explicitly,
\begin{align*}
f_1 = s((1 \, 2 \, 3 \, 4)) &= (1, 0, -1, 1, 0, 1) \\
f_2 = s((1 \, 3 \, 4 \, 2)) &= (-1, 1, 0, 0, -1, 1) \\
f_3 =  s((1\, 4\, 2\, 3)) &= (0, -1, 1, 1, -1, 0).
\end{align*}
Since if we regard $A$ as the $\R^6$ vector $(A_{12},A_{13},A_{14},A_{23},A_{24},A_{34})$, then
$$\langle f_1, A \rangle = A_{12} - A_{14} + A_{23} + A_{34} = A_{12} + A_{23} + A_{34} + A_{41}$$
which is the value of the cycle $(1\, 2\, 3\, 4)$. The computation for $f_2$ and $f_3$ can be checked similarly. The zonotope $Z_\perp$ expressed in this basis is shown in Figure \ref{fig:z4}.
\begin{figure}[ht]
	\centering
		\includegraphics[width=0.85\textwidth]{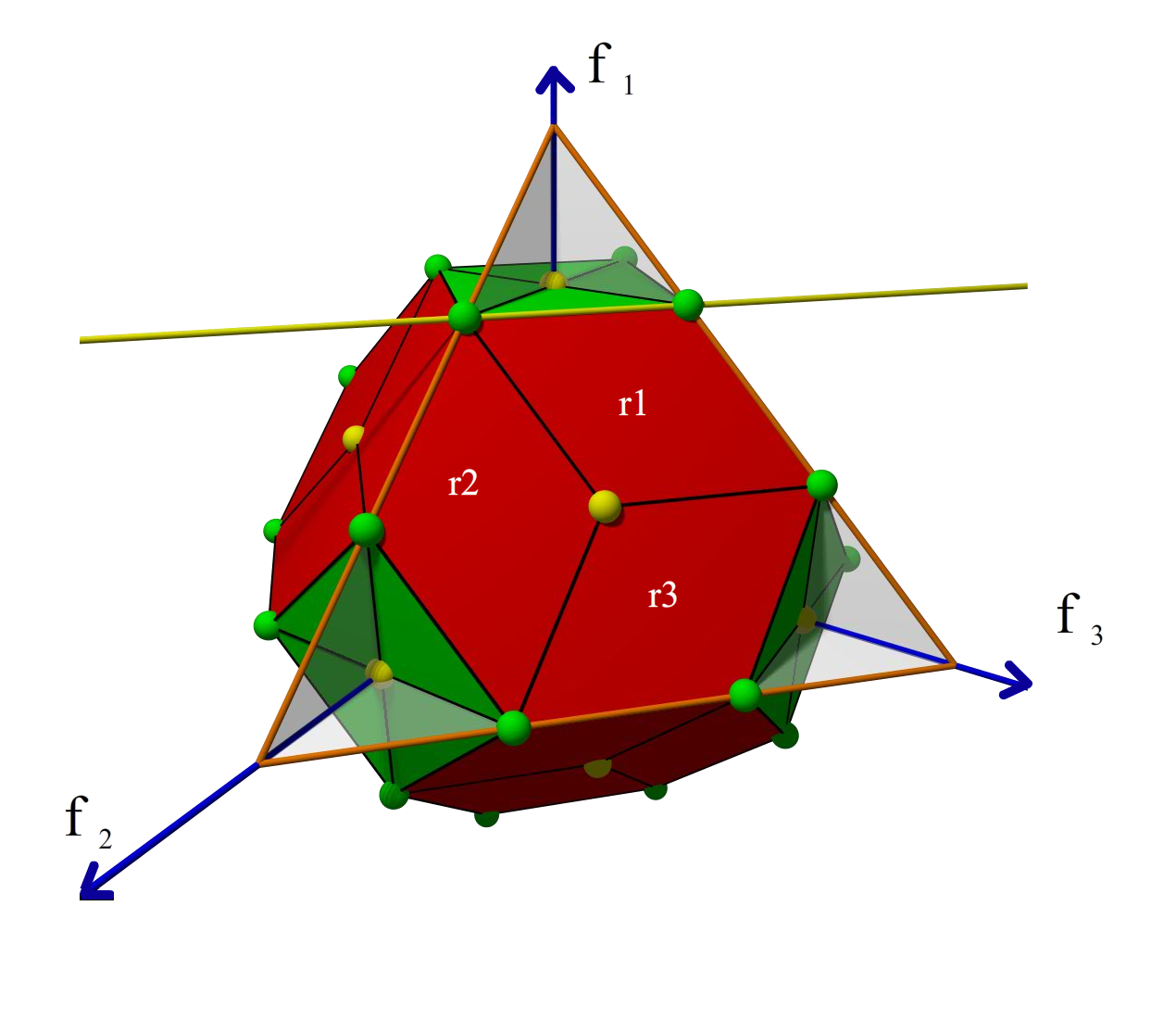}
	\vskip-1cm\caption{$Z_\perp$ with regions colored according to the two cases, up to permutation of the labellings of the items.}
	\label{fig:z4}
\end{figure}
\vskip12pt \noindent
Note that $Z_\perp$ is also a permutahedron; this is due to the complete graph on $4$ elements being self-dual, therefore the graphic and cographic zonotope of Corollary \ref{cor:zonotope} are isomorphic. $Z_\perp$ has 8 hexagonal facets colored red, corresponding to the eight 3-cycles, and 6 square facets colored green, corresponding to the six 4-cycles. The value and critical cycle of the tropical max-plus eigenvalue $\lambda_{max}$ is determined by entirely by the facet of $\lambda_{max} \cdot Z_\perp$ that contains $0$. Each facet is further subdivided into regions corresponding to different formulae for the tropical eigenvector. Three such regions are labeled $r1,r2,r3$ in Figure $\ref{fig:z4}$. This subdivision is due to the fact that the tropical eigenvector is sensitive to permutation of the labeling of the items that fixes the critical cycle. Hence, the square facet is subdivided into 4 regions, while the hexagonal facet is subdivided into 3 regions. Modulo permutation of items, there are only two formulae for the tropical eigenvalue and tropical eigenvector, one for each facet type. To obtain their explicit formulae from Figure ~ \ref{fig:z4}, we only need to know the equations defining boundaries of the facets and their subdivisions. The equations $$\langle f_1,A\rangle = 0, \langle f_2,A\rangle = 0, \langle f_3,A\rangle = 0$$ are three gray planes with orange boundary, defining the three edges shared by adjacent hexagons. The equations 
$$2(\langle f_2,A\rangle+\langle f_3,A\rangle) = \langle f_1,A\rangle, 2(\langle f_1,A\rangle+\langle f_3,A\rangle) = \langle f_2,A\rangle, 2(\langle f_1,A\rangle+\langle f_2,A\rangle) = \langle f_3,A\rangle$$
define the other three edges of the hexagon. Only the edge cut out by the first equation is shown here in yellow. The black lines subdividing the frontal hexagon into three regions are 
$$2\langle f_1, A \rangle = \langle f_2,A\rangle+\langle f_3,A\rangle, 2\langle f_2, A \rangle = \langle f_1,A\rangle + \langle f_3,A\rangle, 2\langle f_3, A \rangle = \langle f_1,A\rangle + \langle f_2,A\rangle.$$
One can compute and see that the frontal hexagon corresponds to the $3$-cycle $(2\, 3\, 4)$, and the three green squares perpendicular to the axes $f_1, f_2, f_3$ correspond to the three $4$-cycles defining these vectors. One can also read off the formula for $\lambda_{max}(A)$ in terms of the cycle values and the inequalities defining the regions, as demonstrated below.
\vskip12pt \noindent
\emph{Formulae for the tropical eigenvalue.}\\
Suppose $\langle f_1, A \rangle, \langle f_2, A \rangle, \langle f_3, A \rangle > 0$. This means $0$ lies in the triangular region bounded by the orange lines. If
\begin{align*}
\langle f_1, A \rangle &\geq 2\cdot(\langle f_2, A \rangle + \langle f_3, A \rangle) \hspace{1em} \mbox{ or } \\
\langle f_2, A \rangle &\geq 2\cdot(\langle f_1, A \rangle + \langle f_3, A \rangle) \hspace{1em} \mbox{ or } \\
\langle f_3, A \rangle &\geq 2\cdot(\langle f_1, A \rangle + \langle f_2, A \rangle),
\end{align*}
which implies that $0$ lies in one of the three small triangles outside of the hexagon, then the 4-cycle corresponding to $f_1$ (or $f_2$, or $f_3$, respectively) is a critical cycle of ~ $A$, and 
$$\lambda_{max}(A) = \frac{1}{4}(\langle f_1, A \rangle) \mbox{ or } \frac{1}{4}(\langle f_2, A \rangle) \mbox{ or } \frac{1}{4}(\langle f_3, A \rangle) \mbox{, respectively}.$$
If
\begin{align*}
\langle f_1, A \rangle &\leq 2\cdot(\langle f_2, A \rangle + \langle f_3, A \rangle) \hspace{1em} \mbox{ and } \\
\langle f_2, A \rangle &\leq 2\cdot(\langle f_1, A \rangle + \langle f_3, A \rangle) \hspace{1em} \mbox{ and } \\
\langle f_3, A \rangle &\leq 2\cdot(\langle f_1, A \rangle + \langle f_2, A \rangle),
\end{align*}
which implies that $0$ lies in the hexagon, then the 3-cycle $(2\, 3\, 4)$ is a critical cycle of ~ $A$, and
$$\lambda_{max}(A) = \frac{1}{6}(\langle f_1, A \rangle + \langle f_2, A \rangle + \langle f_3, A \rangle).$$
\emph{Formulae for the tropical eigenvector.}
The tropical eigenvector can also be read off from $Z_\perp$. Here we give explicit formulae for the red region marked $r1$ and the green region above it. 
\begin{itemize}
	\item The red region $r1$: this region is defined by $\langle f_1,A\rangle > \langle f_2,A\rangle, \langle f_3,A\rangle > 0$, $2(\langle f_2,A\rangle+\langle f_3,A\rangle)>\langle f_1,A\rangle$, $2f_2 < \langle f_1,A\rangle + \langle f_3,A\rangle$, and $2f_1 > \langle f_2,A\rangle + \langle f_3,A\rangle$. In this case,
\[ m(A) = h(A) + \frac{1}{12}\left[ \begin{array}{ccc}
	0 & 0 & 0 \\
	-1 &   5 &   2 \\
	 -2 &   7 &   1 \\
	-3 & 6 & 3	\end{array} \right] \, \left[ \begin{array}{c} \langle f_1,A\rangle \\ \langle f_2,A\rangle \\ \langle f_3,A\rangle \end{array} \right]
	\]
\item The green region above $r1$: this region is defined by $\langle f_1,A\rangle > \langle f_2,A\rangle, \langle f_3,A\rangle > 0$, and $2(\langle f_2,A\rangle+\langle f_3,A\rangle) < \langle f_1,A\rangle$. In this case,
\[ m(A) = h(A) + 
\frac{1}{12}\left[ \begin{array}{ccc}
	0 & 0 & 0 \\
	0 &  3 &  0 \\
	 0 &   3 &  -3 \\
	0 & 0 & -3	\end{array} \right] \, \left[ \begin{array}{c} \langle f_1,A\rangle \\ \langle f_2,A\rangle \\ \langle f_3,A\rangle \end{array} \right]
	\]
\end{itemize}
Formulae for the tropical eigenvalue and eigenvector for other regions can be obtained by permuting the labels of the vertices, corresponding to rotating the zonotope. For example, if one acts on the labellings of the items by $\sigma = (4 \, 2 \, 3)$, then
\begin{align*}
\sigma(\langle f_1,A\rangle) &= \sigma(A_{12} + A_{23} + A_{34} + A_{41}) = A_{13} + A_{34} + A_{42} + A_{21} = \langle f_2, A \rangle
 \\
\sigma(\langle f_2, A \rangle) &= \sigma(A_{13} + A_{34} + A_{42} + A_{21}) = A_{14} + A_{42} + A_{23} + A_{31} = \langle f_3, A \rangle \\
\sigma(\langle f_3, A \rangle) &= \sigma(A_{14} + A_{42} + A_{23} + A_{31}) = A_{12} + A_{23} + A_{34} + A_{41} = \langle f_1, A \rangle.
\end{align*}
Therefore, the action of $\sigma$ on the label of items corresponds to a counter-clockwise 120$^\circ$ rotation of $Z_\perp$ around the center of the hexagonal facet in the picture. The red region $r1$ is mapped to region $r2$, and the formula for the max-plus eigenvector in this region is
\[ m(A) = h(A) + 
	\frac{1}{12}\left[ \begin{array}{ccc}
	0 & 0 & 0 \\
 3 &  -3 &   6\\
2 &  -1 &   5 \\
 1 &  -2  &  7
	\end{array}
	\right] \, \left[ \begin{array}{c} \langle f_1,A\rangle \\ \langle f_2,A\rangle \\ \langle f_3,A\rangle \end{array} \right] \]
\end{ex}
Note that in th example $m(A) - h(A)$ only depends on the $\ST^\perp$ component of $A$. This is true in general, and is linked to the geometric relationship between Tropical Eigenvector and HodgeRank. 
\begin{lem}{}\label{lem:m-h} \noindent
Let $P(A)$ be the $L_2$-projection of $A$ onto $\ST$. Then the tropical eigenvalue of $A - P(A)$ is $\lambda_{max}(A)$, and its max-plus eigenvector is $m(A)-h(A)$.
\end{lem}
\vskip12pt \noindent
\begin{Proof}
Note that $\lambda_{\max}(A)$ is the tropical max-plus eigenvector of $A$ if and only if for all $i$, $\ds\max_j\left(A_{ij} - (m_i - m_j)\right) =~ \lambda_{\max}(A)$. Since $P(A)_{ij} = h_i - h_j$, we have
\begin{align*} 
\lambda_{\max}(A) &= \max_j\left\{(A - P(A))_{ij} + (h_i-h_j)-(m_i-m_j)\right\} \hspace{1em} \mbox{for all $i$} \\
&= \max_j \left\{(A - P(A))_{ij} - [(m-h)_i-(m-h)_j] \right\} \hspace{1em} \mbox{for all $i$.}
\end{align*}
Hence $m(A)-h(A)$ is the max-plus eigenvector of $A-P(A)$.
\end{Proof}
\subsection{Proof of Lemma \ref{lem:list.prop}}\label{subsec:proof.lem}
Part (1) follows from a direct computation. Statements on the HodgeRank vector follows from Lemma \ref{lem:h}. We shall prove the stated results for the tropical eigenvector and principal eigenvector. \\
Since the entries of the tropical eigenvector are piecewise linear in terms of the entries of $A$, $m(cA) = cm(A)$. If $A$ is strongly transitive, then the $\ST^\perp$ component of $A$ is zero, hence $m(A + A') - h(A + A') = m(A') - h(A')$ by Lemma \ref{lem:m-h}. But $h(A + A') = h(A) + h(A') = m(A) + h(A')$ since $A \in \ST$, hence $m(A + A') = m(A) + m(A')$. The corresponding statements for $X \in \X$ ~ follow. \\
Example \ref{exam:pev} in Section \ref{sec:proof} shows that the Principal Eigenvector does not have property (2). To prove (3), suppose the matrix $X'$ is strongly transitive with score vector $s$. Then $\ds(X \circ X')_{ij} = X_{ij}s_i/s_j$. Multiplying out, we see that the vector $vs$ with $(vs)_i = v_is_i$ is an eigenvector of $X \circ X'$ with positive entries, hence it is the principal eigenvector by Perron-Frobenius theorem. \qed
\section{Proof of Theorem \ref{thm}}\label{sec:proof}
\begin{lem}{}\label{lem:find.one}
(1) \hspace{1em} To prove Theorem 1 for HodgeRank and Tropical Eigenvector, it is sufficient to find one matrix $A \in \A$ such that $h(A) = (0,\ldots,0)$, and $m(A)$ induces a ranking without ties.\\
(2) \hspace{1em} To prove Theorem 1 for Tropical and Principal Eigenvector, it is sufficient to find one matrix $X \in \X$ such that $m(X) = (1, \ldots, 1)$ and $v(X)$ is a ranking without ties. 
\end{lem}
\begin{Proof}
For (1): suppose $A$ is such a matrix. Fix a ranking $\sigma$. By invariance under relabeling of items, it is sufficient to find a matrix $B$ such that $r(m(B)) = r(m(A))$ and $r(h(B)) = \sigma$. Consider 
$$B := A + \epsilon \cdot W \mbox{ for some }\epsilon > 0, W \in \ST \mbox{ such that }r(h(W)) = r(m(W)) = \sigma $$
By Lemma ~ \ref{lem:list.prop}, $m(B) = m(A) + \epsilon \cdot m(W)$, hence for $\epsilon$ sufficiently small $r(m(B)) = r(m(A))$, while $r(h(B)) = r(h(W)) = \sigma$. \qed
\vskip12pt \noindent
For (2): suppose $X$ is such a matrix. Let $M \in \X$ be a strongly transitive matrix with $r(v(M)) = r(m(M)) = \sigma$. For $k > 0$, let $Y := X \circ M^{(k)}$. By Lemma \ref{lem:list.prop}, $m(Y) = m(X) \circ (m(M)^{k})$, so $r(m(Y)) = r(m(M)) = \sigma$. It would be sufficient to show that one can choose $k$ small enough such that $r(v(Y)) = r(v(X))$. \\
Indeed, from classical results in linear algebra (see, for example, \cite{alg}), the largest eigenvalue of $X$ is a continuous function in its entries, hence is also a continuous function in its upper-diagonal entries when $X \in \X$. Furthermore, since $\lambda$ has multiplicity 1,  any column of the adjugate matrix $\adj(X - \lambda \textbf{I})$ is a principal eigenvector, where $\textbf{I}$ is the identity matrix. Since each entry of this matrix is a cofactor of an entry in $X - \lambda \textbf{I}$, each $v_i$ is a polynomial in $\lambda$ and entries of $X$. Let $\ds \epsilon = \min_{i,j}|v_i(X) - v_j(X)|$. Since there are only finitely many $v_i$'s, one can find a $\delta > 0$ such that the ordering of the $v_i$'s are preserved when each upper-diagonal entry of $X$ is perturbed by $e^\delta$. Hence $\ds k = \delta/\log(\|M\|_\infty)$ would do. 
\end{Proof}
\textbf{Proof of Theorem \ref{thm}} \\
\textbf{For HodgeRank and Tropical Eigenvector} 

We shall construct the matrix $A$ in Lemma \ref{lem:find.one} as follows: start with a matrix $A'$ such that
\begin{itemize}
	\item (1) $h(A') = 0$.
	\item (2) $A'_{12} > A'_{23} > \ldots > A'_{(n-1) \, n} > A'_{n1} > 0$.
	\item (3) For $\mu := \ds \frac{A'_{12} + A'_{23} + \ldots + A'_{(n-1)\, n} + A'_{n\, 1}}{n}$, we have $A'_{i \, (i+1)} \neq \mu$ for all $i$.
\end{itemize}
Such a matrix $A'$ exists since $\dim(\A) > n$ for all $n > 3$. Let $A := A' + k \cdot B$, where $B \in \A$ is the matrix with upper diagonal entries
\[ B_{i,{i+1}} = 1 \mbox{ for all }i \leq n-1, \hspace{0.5em} B_{1,n} = -1, \hspace{0.5em} B_{ij} = 0 \mbox{ else}.
\]
Note that $A$ satisfies properties (1) to (3), and for sufficiently large $k$, the following also hold.
\begin{itemize}
	\item (4) For each row $i$, $\max_jA_{ij} = A_{i\, (i+1)}$.
	\item (5) $\lambda_{\max} = \mu$. That is, the $n$-cycle $1 \to 2 \to \ldots \to n \to 1$ is a critical cycle of $A$.
\end{itemize}
By Corollary \ref{cor:gaubert}, property (5) implies that the tropical eigenvector $m$ in this case is unique. Property (4) implies $m_i - m_{i+1} = A_{i \, (i+1)} - \lambda$ for all $i$.
By (2) and (3), there exists $j \in \{1, \ldots, n\}$ such that $A_{i \, (i+1)} > \lambda$ for all $i < j$, and $A_{i \, (i+1)} < \lambda$ for all $i \geq j$. This implies
\[ m_j < m_{j-1} < \ldots < m_1 < m_n < \ldots < m_{j+1}. \]
Hence $A$ is the matrix needed. \qed \\
\textbf{For HodgeRank and Principal Eigenvector}

By Corollary \ref{cor:gaubert}, this result follows immediately from the above by a limit argument. Specifically, let $X$ be a matrix that induces the ranking $(\sigma_1,\sigma_2)$ on the pair HodgeRank and Tropical Eigenvector. Let $\ds \epsilon := \min_{i,j}|m(X)_i - m(X)_j|$. By Corollary \ref{cor:gaubert}, one can choose sufficiently large $k$ such that the matrix $Y := X^{(k)}$ satisfies $|v_i(Y) - m_i(X)| < \epsilon/3$ for all $i$. So $r(v(Y)) = r(m(X)) = \sigma_2$, and by Lemma \ref{lem:list.prop}, $r(h(Y)) = \sigma_1$. Hence $Y$ is a matrix that realizes the pair $(\sigma_1,\sigma_2)$ for HodgeRank and Principal Eigenvector. \qed 
\vskip12pt \noindent
\textbf{For Tropical and Principal Eigenvector}

To find the desirable $X$ in Lemma \ref{lem:find.one}, we consider a family of matrices called \emph{perturbed comparison matrices}, first introduced by Astuti and Garnadi \cite{indo} and a year later by Farkas \cite{farkas}. These are strongly transitive MPCMs with the first row and column perturbed by a noise vector. Explicit formulae for the principal eigenvalue and eigenvector are given in \cite{farkas,indo}. 

For our case, fix a number $L > 1, L \in \mathbb{Q}$. Define $s \in \R^n$ by
\[ s_i = 1 \mbox{ for all }i < n, \,\, s_n = \frac{1}{L}.  \]
Let the upper-diagonal entries of $X \in \X$ be
\[ \ds X_{ij} = \frac{s_i}{s_j} \mbox{ if } i \neq 1, \hspace{1em} X_{1j} = \frac{1}{s_j} \cdot \delta_j \]
where $\delta = (\delta_2, \ldots, \delta_n) \in \mathbb{Q}^{n-1}$  is the noise vector, with 
\begin{equation}
\delta_2 < \delta_3 < \ldots < \delta_{n-1} = L, \mbox{ and } \delta_n = \frac{1}{L^2} \label{eqn:delta}
\end{equation}
Then
\[ X = \left[ 
\begin{array}{cccccc}
1 & \delta_2 & \delta_3 & \cdots & L & \frac{1}{L} \\
\frac{1}{\delta_2} & 1 & 1 & \cdots & 1 & L \\
\frac{1}{\delta_3}& 1 & 1 & \cdots & 1 & L \\
\vdots & \vdots & \vdots & \vdots & \vdots & \vdots \\
\frac{1}{L} & 1 & 1 & \cdots & 1 & L \\
L & \frac{1}{L} & \frac{1}{L} & \cdots & \frac{1}{L} & 1
\end{array}
\right]. \]
By \cite{indo}, the principal eigenvector $v$ of $X$ is
\[ v = \alpha_1 e_1 + \alpha_2 s + \alpha_3 w \]
where 
\[ e_1 = \left[\begin{array}{c} 1\\ 0\\ 0\\ \vdots\\ 0 \\ 0\end{array}\right],  s = \left[\begin{array}{c} 1\\ 1\\ 1\\ \vdots\\ 1 \\ \frac{1}{L}\end{array}\right], w = \left[\begin{array}{c} 0\\ \frac{1}{\delta_2}-1\\ \frac{1}{\delta_3}-1\\ \vdots\\ \frac{1}{L}-1 \\ \frac{1}{L}\left(\frac{1}{L^2}-1\right)\end{array}\right] \]
and $\alpha := (\alpha_1,\alpha_2,\alpha_3)$ is a column vector of the adjugate matrix $\adj(r\textbf{I} - Z)$, where $\textbf{I}$ is the identity matrix, and
\[ Z := \left[\begin{array}{ccc}
0 & a & b \\ 1 & n & c \\ 1 & 1 & 0
\end{array}\right], \]
with $$\ds a = \sum_{i=2}^n(\delta_i-1), b = \sum_{i=2}^n(\delta_i -1)\left(\frac{1}{\delta_i}-1\right), c = \sum_{i=2}^n\left(\frac{1}{\delta_i}-1\right),$$ 
and $r$ is the principal eigenvalue of $Z$, that is, it is the root of maximal modulus of the characteristic polynomial
\begin{equation} p(t) = t^2(t-n) + b(n-1) - ac. \label{eqn:char} \end{equation}
Since $L > 1$ is the maximal element in each row, $m(X) = (1, \ldots, 1)$. By Lemma \ref{lem:find.one}, it is sufficient to prove that there exists choices of $\delta$ such that $v(X)$ induces a complete ranking. In fact, we claim that any rational $\delta$ satisfying Equation (\ref{eqn:delta}) would do. Indeed, computing an explicit formula for $\alpha$, we obtain
\[ \alpha_1 = (r-n)r - c, \hspace{0.5em} \alpha_2 = r + c, \hspace{0.5em} \alpha_3 = r - n +1. \]
Then, 
\begin{align*}
v_1 &= (r-n)r -c + r + c = (r-n+1)r \\
v_i &= r+c + (r-n+1)\cdot \left(\frac{1}{\delta_i} - 1\right) \mbox{ for }2 \leq i < n-1 \\
v_{n-1} &= r+c + (r-n+1)\cdot \left(\frac{1}{L} - 1\right) \\
v_n &=(r+c) \cdot \frac{1}{L} + (r-n+1)\cdot\frac{1}{L}\left(\frac{1}{L^2}-1\right)
\end{align*}
Note that $v_1$ is a quadratic in $r$, while $v_i$ are linear for $i > 1$. Since the $\delta_i$'s are distinct, $v_i \neq v_j$ for all $2 \leq i, j \leq n-1$. 
 Suppose $v_1 = v_i$ for some $i \neq 1$, or $v_n = v_i$ for some $i \neq n$. In the first case, $r$ is a root of a quadratic with coefficients in $\mathbb{Q}$. In the second case, $r$ is a root of a linear equation with coefficients in $\mathbb{Q}$. Either cases imply the cubic $p$ in Equation (\ref{eqn:char}) must be factorisable into a quadratic times a linear term over $\mathbb{Q}$. That is
\[ p(t) = t^3 - nt^2 + b(n-1) - ac = (t^2 + et + 1)(t-g) \mbox{ for some }e,g \in \mathbb{Q} \]
Equating coefficients, we have $eg = 1$, $e-g = -n$, and $b(n-1)-ac = -g$. The first two equations imply $1/g - g + n = 0$, hence $g = (n - \sqrt{n^2+4})/2$ or $(n + \sqrt{n^2+4})/2$. But $g \in \mathbb{Q}$, therefore $n^2 + 2^2$ must be a square natural number, since $n \in \mathbb{N}$. However, there is no positive Pythagorean triple involving 2, hence no such $g$ exists. Therefore, the $v_i$'s are all distinct. \qed
\subsection{Realizability of ranking triples}
It is natural to ask whether all ranking triples can be realized. For $n = 3$, the answer is no.
\begin{cor}
For $n = 3$, if $X \in \X$, then $h(X) = m(X) = v(X)$ (up to scaling). That is, all methods produce the same score vector, and hence the same ranking. 
\end{cor}
The result is a direct computation: if $X \in \X$ is not strongly transitive, there can be only one maximal cycle of length 3, hence the tropical eigenvector is always unique. In this case, one can plug in the formula for the HodgeRank vector (as row geometric mean) to verify that it is indeed also the tropical and principal eigenvector of $X$. We can also see this geometrically in $\A$ for the pair HodgeRank and Tropical Eigenvector: in this case $\ST^\perp$, viewed as vectors of the upper-diagonal entry, is spanned by the vector $(1,-1,1)$. Hence the standard 3-cube centered at $A$ will always hits $\ST$ at either the corner $A + \lambda_{\max}\cdot (1,-1,1)$, or the opposite corner $A + \lambda_{\max}\cdot (-1,1,-1)$, therefore $h(A) = m(A)$ for all $A \in \A$.
\vskip12pt \noindent
The case $n > 3$ is substantially more difficult. If one thinks of each matrix $X \in \X$ as being part of a family $\{X^{(k)}: k > 0\}$, then Corollary \ref{cor:gaubert} implies that Principal and Tropical Eigenvector induce different rankings only on a finite interval. In general, we do not about the behavior of the principal eigenvector as $k \to 0$, and the transition in induced ranking as $k \to \infty$.  Interestingly, simulation results suggest the following.
\begin{conj}
As $k \to 0$, the induced ranking of Principal Eigenvector converges to that induced by HodgeRank.
\end{conj} 
Since components of the normalized principal eigenvector vary smoothly with $k$, the conjecture suggests that not all ranking regions may be realizable, especially when the ranking induced by HodgeRank and Tropical Eigenvector agree, since in this case the ranking induced by Principal Eigenvector is `sandwiched' between these two. This is illustrated in the example below.
\begin{ex}\label{exam:pev}
Here our matrix $X \in \X$ is
\[X = \left[\begin{array}{cccc}
1 & 1.57 & 0.72 & 0.70 \\
0.63 & 1 & 1.52 & 0.65 \\
1.38 & 0.65 & 1 &1.57 \\
1.45 & 1.52 & 0.63& 1
\end{array}\right] \]
The corresponding vectors, normalized to have the first component be 1 and rounded in 3 decimal places, are
\begin{align*}
v(X) = \left[\begin{array}{cccc}
1 & 0.991 & 1.191 & 1.151
\end{array}\right] &\Rightarrow \mbox{ranking: } 3 > 4 > 1 > 2\\
h(X) = \left[\begin{array}{cccc}
1 & 0.942 & 1.155 & 1.151
\end{array}\right] &\Rightarrow \mbox{ranking: } 3 > 4 > 1 > 2 \\
m(X) = \left[ \begin{array}{cccc} 
1 & 0.979 & 0.989 & 0.968
\end{array} \right] &\Rightarrow \mbox{ranking: } 1 > 3 > 2 > 4
\end{align*}
One may note that both the matrix $X$ and the corresponding scores do not have `crazy' entries that can intuitively indicate intransitivity. For example, if one uses the consistency index $(\lambda(X)-n)/(n-1)$ suggested by \cite{fuzzy}, where $\lambda(X)$ is the principal eigenvalue of $X$, then the consistency index for this case is 0.07073, well within the proposed $0.1$ recommended cut-off. This indicates that the consistency index may not be a good measure of ranking agreement between methods.
\begin{figure}[ht]
	\centering
		\includegraphics[width=0.85\textwidth]{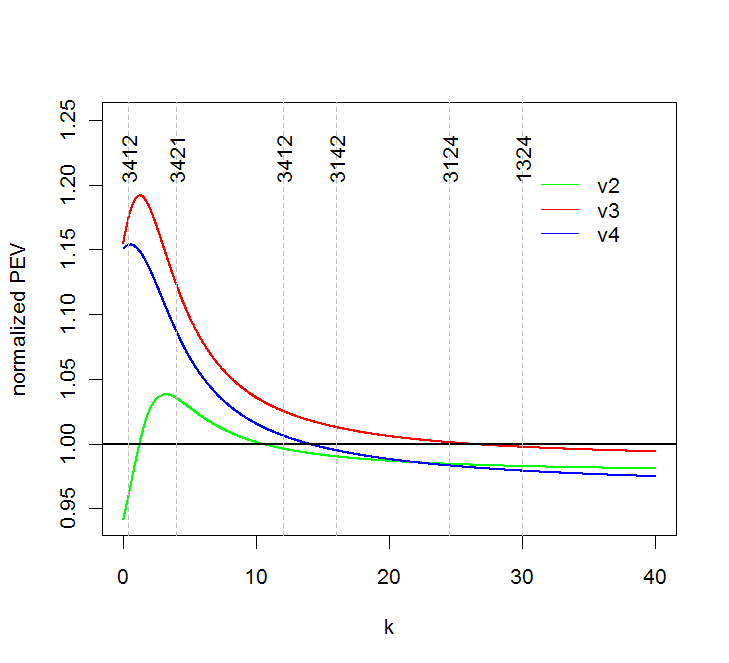}
	\vskip-0.5cm \caption{Behavior of components of the normalized principal eigenvector of $X^{(k)}$ over a range of $k$, and the corresponding induced rankings at various points.}
	\label{fig:pevplot}
\end{figure}
\end{ex}
Figure \ref{fig:pevplot} is a plot of the components of the principal eigenvector of $X^{(k)}$ over a range of $k$, normalized so $v_1(X^{(k)}) = 1$. This is represented by the black line. The green, red and blue lines are $v_2(X^{(k)}), v_3(X^{(k)})$ and $v_4(X^{(k)})$, respectively. Note that each $v_i$ is a smooth function in $k$. A change in the induced ranking happens when one curve crosses over another. The sequence of induced rankings (in decreasing order) is shown at various marked points on the plot. The ranking of Principal Eigenvector agrees with that of Tropical Eigenvector after $k = 30$.

Examples like this suggest that not all ranking triples may be realizable. Studying this problem would give further information on the relation between the three methods.
\section{Discussions and Open Problems}\label{sec:discuss}
While we have completely understood the picture for HodgeRank v.s. Tropical Eigenvector, the behavior of the pair Principal and Tropical Eigenvector is still unclear. Our current construction of matrices realizing a particular ranking pair for these two is somewhat artificial. Simulations suggested that one often hit a region of non-trivial ranking pair far more often than hitting a perturbed comparison matrix. Recently, Ottaviani and Sturmfels \cite{bernd} studied characterizations of matrices with eigenvectors lie in a given subspace, called the Kalman variety. To completely answer the question for the pair Principal and Tropical Eigenvector, one needs a tropical Kalman variety, and study how it intersects with the usual Kalman variety.
\vskip12pt \noindent
A directly related question of interest is the following. \\
\textbf{Problem 1:} \emph{For `reasonable' probability distributions on $\epsilon$, what is the probability of $A$ falling into a region in which some (or all three methods) give different rankings?}
\vskip12pt 
What constitutes as `reasonable' depends on the problem's context and ease of computation. For example, if one considers a uniform distribution on $\ST^\perp$, then the question reduces to computing the relative volumes of the regions of disagreement. We note that these regions change depending on the true score vector $s$, corresponding to a `signal-to-noise' behavior: if $s$ is close to the origin, indicating that the rater does not have a strong preference between items, disagreement and false rankings easily occur when the noise is large. 
\vskip12pt \noindent
\textbf{Problem 2:} \emph{What happens when $A$ has missing entries?}
\vskip12pt 
Missing entries do occur in practice. For example, it is a known issue in the Analytic Hierarchy Process, since raters may be unwilling to perform all ${n \choose 2}$ pairwise comparisons for large $n$. A number of papers have been devoted to this problem in the literature, mainly in the context of HodgeRank and Principal Eigenvector \cite{lekheng, harker,bozoki}. 

So far the main approach to this problem is interpolation: one attempts to fill out the missing entries by maximizing the some measure of consistency of the resulting matrix \cite{bozoki}. From the viewpoint of graph theory, missing entries imply we have a subgraph of the complete graph on $n$ vertices. \cite{lekheng} proved that if all three-cycles of the corresponding subgraph have value 0, then the matrix has a rank-1 completion. It may be interesting to look at low-rank completion of AHP matrices, as suggested in the work of \cite{gleich}. 
\vskip12pt \noindent
A closely related question to both problem 1 and 2 is the following.\\
\textbf{Problem 3:} \emph{What happens when the number of voters tend to $\infty$, or the number of items tend to $\infty$?}
\vskip12pt 
In some sense this is a question on a different scale. Here one would be less worried about small differences in ranking (say, a difference by one transposition), but more on how the methods compare in terms of statistical optimality and computational efficiency. By studying the behavior of these methods in smaller dimensions, one can gain insights on this problem.
\section{Summary}\label{sec:end}
In this paper, we considered the problem of obtaining a cardinal ranking from an $n \times n$ pairwise comparison matrix on additive and multiplicative scales. We studied the mathematical properties and connections of three proposed methods: Principal Eigenvector, HodgeRank and Tropical Eigenvector. We noted that Tropical Eigenvector can be thought of as a version of Principal Eigenvector on additive paired comparison matrices. Our main theorem states that for $n \geq 4$, all regions of ranking differences can be realized between any pair of these three methods. This indicates that the choice of method in practice deserve further study. We interpreted this result in the ranking context and listed some future research directions. 
\section{Acknowledgment}
The topic of this paper was formulated after the workshop `Mathematics of Ranking' at the American Institute of Mathematics in August 2010, organized by Shivani Agarwal and Lek-Heng Lim. The author wishes to thank Lek-Heng Lim and Bernd Sturmfels for many helpful discussion and advise. Research supported by the Berkeley Optimization Initiative sponsored by Jeffrey Bohn. NMT is a Vietnam Education Foundation (VEF) Fellow. The opinions, findings and conclusions stated herein are those of the author and do not necessarily reflect those of VEF.
\bibliographystyle{elsarticle-num}
\bibliography{references}

\begin{thebibliography}{10}
\expandafter\ifx\csname url\endcsname\relax
  \def\url#1{\texttt{#1}}\fi
\expandafter\ifx\csname urlprefix\endcsname\relax\def\urlprefix{URL }\fi
\expandafter\ifx\csname href\endcsname\relax
  \def\href#1#2{#2} \def\path#1{#1}\fi

\bibitem{lekheng}
X.~Jiang, L.-H. Lim, Y.~Yao, Y.~Ye, Statistical ranking and combinatorial hodge
  theory, Mathematical Programming 127 (2011) 203--244,
  10.1007/s10107-010-0419-x.

\bibitem{elsner}
L.~Elsner, P.~van~den Driessche, Max-algebra and pairwise comparison matrices,
  Linear Algebra and its Applications 385 (2004) 47 -- 62.
\newblock \href {http://dx.doi.org/DOI: 10.1016/S0024-3795(03)00476-2}
  {\path{doi:DOI: 10.1016/S0024-3795(03)00476-2}}.

\bibitem{farkas}
A.~Farkas, The analysis of the principal eigenvector of pairwise comparison
  matrices, Manuscript (2010).

\bibitem{fuzzy}
T.~L. Saaty, The {A}nalytic {H}ierarchy {P}rocess, McGrawHill, New York, 1980.

\bibitem{saari}
D.~A. Saari, A new way to analyze paired comparison rules, Manuscript (2010).

\bibitem{crawford}
G.~B. Crawford, The geometric mean procedure for estimating the scale of a
  judgement matrix, Mathematical Modelling 9 (1987) 327--334.

\bibitem{dong}
Y.~Dong, G.~Zhang, W.~Hong, Y.~Xu, Consensus models for {AHP} group decision
  making under row geometric mean prioritization method, Decision Support
  Systems 49 (2010) 281 -- 289.

\bibitem{bernd.trop}
D.~Maclagan, B.~Sturmfels, Introduction to {T}ropical {G}eometry, Manuscript
  (2009).

\bibitem{butkovic}
P.~Butkovi\v{c}, Max-linear Systems: Theory and Algorithms, Springer, 2010.

\bibitem{bapat}
R.~Bapat, A max version of the {P}erron-{F}robenius theorem, Linear Algebra and
  Its Applications 275 - 276 (1997) 3 -- 18.

\bibitem{gaubert}
M.~Akian, R.~Bapat, S.~Gaubert, Asymptotics of the {P}erron eigenvalue and
  eigenvector using max-algebra, Comptes Rendus de l'Académie des Sciences -
  Series I - Mathematics 327~(11) (1998) 927 -- 932.
\newblock \href {http://dx.doi.org/DOI: 10.1016/S0764-4442(99)80137-2}
  {\path{doi:DOI: 10.1016/S0764-4442(99)80137-2}}.

\bibitem{elsner10}
L.~Elsner, P.~van~den Driessche, Max-algebra and pairwise comparison matrices,
  ii, Linear Algebra and its Applications 432~(4) (2010) 927 -- 935.
\newblock \href {http://dx.doi.org/DOI: 10.1016/j.laa.2009.10.005}
  {\path{doi:DOI: 10.1016/j.laa.2009.10.005}}.

\bibitem{ziegler}
G.~M. Ziegler, Lectures on Polytopes, Springer-Verlag, 1995.

\bibitem{alg}
J.~H. Wilkinson, Algebraic Eigenvalue Problem, Clarendon Press, Oxford, 1965.

\bibitem{indo}
P.~Astuti, A.~D. Garnadi, On eigenvalues and eigenvectors of perturbed
  comparison matrices, ITB Journal of Science 41 A (2009) 69 -- 77.

\bibitem{bernd}
G.~Ottaviani, B.~Sturmfels, Matrices with eigenvectors in a given subspace,
  arXiv:1012.1016v1 (2010).

\bibitem{harker}
P.~T. Harker, Incomplete pairwise comparisons in the analytic hierarchy
  process, Mathematical Modelling 9~(11) (1987) 837 -- 848.

\bibitem{bozoki}
S.~Boz\'oki, F.~F\"ul\"op, L.~R\'onyai, On optimal completion of incomplete
  pairwise comparison matrices, Mathematical and Computer Modelling 52 (2010)
  318 -- 333.

\bibitem{gleich}
D.~Gleich, L.-H. Lim, Rank aggregation via nuclear norm minimization,
  Manuscript (2010).

\end{thebibliography}
\end{document}